\documentstyle[seceq,preprint]{jpsj}
\newcommand{\ddp}{{\rm d}p}
\newcommand{\dk}{{\rm d}k}
\newcommand{\intinf}[1]{\int_{-\infty}^{\infty} #1}

\newcommand{\tot}{\rm tot}
\newcommand{\ph}{\rm ph}

\newcommand{\eps}{\epsilon}
\newcommand{\al}{\alpha}

\newcommand{\wup}{w_{\uparrow}}
\newcommand{\wdn}{w_{\downarrow}}
\newcommand{\up}{\uparrow}
\newcommand{\dn}{\downarrow}
\newcommand{\Piup}{\Pi_{\uparrow}}

\newcommand{\wB}{w_{\rm B}}

\newcommand{\ws}{w_{\rm s}}
\newcommand{\wsigma}{w_{\sigma}}
\newcommand{\wc}{w_{\rm c}}
\newcommand{\wh}{w_{\rm h}}

\newcommand{\wF}{w_{\rm F}}

\newcommand{\wFl}{w_{{\rm F},1}}
\newcommand{\wFs}{w_{{\rm F},\sigma}}
\newcommand{\wFsp}{w_{{\rm F},\sigma '}}

\newcommand{\rhoc}{\rho_{\rm c}}

\newcommand{\B}{\rm B}

\newcommand{\PiF}{\Pi_{1}}

\newcommand{\Pis}{\Pi_{\rm s}}
\newcommand{\Pic}{\Pi_{\rm c}}
\newcommand{\Pisigma}{\Pi_{\sigma}}

\def\runtitle{ Fractional Exclusion Statistics for the {\it t-J}  Model}
\def\runauthor{Yusuke {\sc Kato} and Yoshio {\sc Kuramoto}}
\title
{
Fractional Exclusion Statistics for the {\it t-J}  Model \\
with Long Range Exchange and Hopping
} 
\author
{
Yusuke {\sc KATO}\footnote{e-mail address: kato@cmpt01.phys.tohoku.ac.jp} and Yoshio {\sc KURAMOTO}\footnote{e-mail address: kuramoto@cmpt01.phys.tohoku.ac.jp}
}
\inst
{
Department of Physics, Tohoku University, Sendai 980-77
}
\recdate{
\hspace{5cm}}
\abst
{
We construct thermodynamics of the one-dimensional supersymmetric {\it t-J}  model with the $ 1/\sin^2$ interaction and hopping.  The thermodynamics is described exactly in terms of free spinons and holons obeying Haldane's fractional exclusion statistics at all temperatures. 
Moreover, at low temperatures the semionic spinons and holons decouple 
resulting in the spin-charge separation in thermodynamic properties.
We obtain explicit results for the spin and charge susceptibilities and specific heat, and interpret them in terms of the fractional exclusion statistics.  
Extension to the multi-component {\it t-J}  model shows that the excitations obey either fractional statistics for g-ons with partial polarization of components, or the parafermionic one without polarization.
}
\kword
{
{\it t-J}  model, Sutherland model, thermodynamics, 
fractional exclusion statistics, g-on, parafermion
}

\begin{document}
\sloppy
\maketitle
\clearpage
\section{Introduction}
Recently, the Calogero-Sutherland model \cite{Suth} has attracted much interest among many physicists. The dynamical symmetry of the system enables us to calculate the energy spectrum, explicit form of the wave function, and static and dynamical correlation functions. From explicit results on these quantities, the system has been identified with a free-particle system  governed by the fractional exclusion statistics proposed by Haldane. \cite{Haldanefrac} 

Among the family of the Calogero-Sutherland model, the lattice models such as the Haldane-Shastry model\cite{Haldane1,Shastry} and supersymmetric (SUSY) {\it t-J}  model \cite{KY} play important roles in the condensed matter physics. These lattice models have the Gutzwiller-Jastrow type ground state wavefunction. \cite{KY} In the ground state energy of the {\it t-J}  model, magnetic and charge parts decouple from each other. This has been interpreted as the complete spin charge separation. \cite{KY} In the study of the dynamics at $T=0$, Ha and Haldane identified  holon, spinon, and antiholon as elementary excitations. \cite{Ha} From these results, we regard the {\it t-J}  model as the fixed-point model realizing the ideal Tomonaga-Luttinger liquid.  

In this paper, we clarify the relation between the {\it t-J}  model and the fractional exclusion statistics from the viewpoint of thermodynamics. For the thermodynamics of {\it t-J} model, an approach has been proposed by Wang, Liu, and Coleman\cite{Wang} and another by Ha and Haldane. \cite{squeeze} These approaches, however, rely on empirical assumptions on the state-counting rule.  In this paper we construct the thermodynamics of the {\it t-J}  model exactly by using the mapping from the continuum Sutherland model in the strong coupling limit. \cite{Poly2}   

In the next section, we construct the thermodynamics of the SU($K$, $1$) {\it t-J} model by using the method of Sutherland and Shastry. \cite{S2} In this method, we can formulate thermodynamics of the SU($K$,1) model for arbitrary value of $K$. 

In \S 3-5, we consider the SU(2,1) model. In \S 3, we show that the {\it t-J} model at all temperatures is described in terms of free particles obeying the fractional exclusion statistics. \cite{Haldanefrac} In \S 4, we deal with the low temperature range and reduce the description of the multicomponent statistics to so-called g-ons. In \S 5, we present explicit results on the spin and charge susceptibilities and the specific heat.  These results are interpreted from the viewpoint of the g-ons obtained in \S 4. 

In \S 6, we extend the low temperature description to the general SU($K$,1) symmetry. We show that the g-on description works well when all the chemical potentials for $K$ species are different.  On the other hand, if all species share the same value of chemical potential, the thermodynamics are described in terms of free parafermions instead of g-ons. 

A shorter account of our results has been presented elsewhere. \cite{KuramotoKato} In this paper, we explain the calculation in detail and present additional analytical and numerical results. 
\section{Formulation}
In this section, we construct thermodynamics of the SU($K$,1) {\it t-J} model with $1/\sin^2$ interaction. 
First we explain the relation between the continuous SU($K$, 1) Sutherland model in the strong coupling limit and the SU($K$,1) {\it t-J} model. Next we construct thermodynamics of the SU($K$,1) {\it t-J} model, by using the knowledge of the energy spectrum of the SU($K$,1) Sutherland model. Last we rewrite thermodynamic quantities in a simpler form, by introducing a new variable. 

We start from the SU$(K,1)$ Sutherland model for $N$ particle systems: 
\begin{equation}
{\cal H_{\lambda}}=
         -\frac12\sum_{i=1}^{N}\frac{\partial ^2}{\partial x_i^2}
         +\frac{\pi^2}{L^2}\sum_{i<j}
         \frac{\lambda(\lambda - \tilde P_{ij})}{\sin^2
[\pi\left(x_{i}-x_{j}\right)/L]},
\label{SP}
\end{equation}
where $x_{i}$ is the coordinate of the $i$-th particle, $L$ is the linear dimension of the system, and $\lambda$ is a coupling parameter. 
$\tilde P_{ij}$ is given in terms of the exchange operator $P_{ij}$ for the internal degrees of freedom of particles by
\begin{equation}
\tilde P_{ij}\equiv \left\{
\begin{array}{cc}
-P_{ij}&\quad \mbox{ if both $i$ and $j$ are fermions}\\
 P_{ij}&\quad \mbox{otherwise}. 
\end{array}\right.
\end{equation}
The energy spectrum $E_{\lambda}$ of (\ref{SP}) has been obtained \cite{S2,Bernard,Kato} as follows: 
\begin{equation}
E_{\lambda}=\sum_{\kappa=-\infty}^{\infty}\frac{\kappa^2 \nu(\kappa)}{2}+\frac{\pi^2 \lambda}{L^2} \sum_{\kappa=\infty}^{\infty}\sum_{\kappa'=-\infty}^{\infty}\left|\kappa -\kappa '\right|\nu (\kappa )\nu (\kappa' )+E_{0}.
\label{Elambda}
\end{equation}
Here $\kappa $ runs over integers describing the momemtum. $\nu (\kappa)$ is given by $\nu_{\B}(\kappa)+\sum_{\sigma=1}^{K}\nu_{\sigma}(\kappa )$; $\nu_{\B}(\kappa)(=0,1,2\cdots)$ and $\nu_{\sigma }(\kappa)(=0,1)$ are the momentum distribution functions of bosons and fermions with species index $\sigma$ (henceforth referred to as spin), respectively. $E_{0}$ is given by
\begin{equation}
E_{0}=\frac{\pi^2\lambda^2 N\left(N^2-1\right)}{6L^2}. 
\end{equation}

Let us consider the strong coupling limit $\lambda \rightarrow \infty$ in order to relate the Sutherland model with the {\it t-J} model. In this limit, particles localize with a lattice spacing $L/N$. Up to ${\cal O}(\lambda)$, there are two kinds of degrees of freedom; one is the vibration around the lattice points and the other is the exchange of particle species between the two lattice points. The former corresponds to a free phonon while the latter does to a hopping of an electron or exchange of spins. For the Hamiltonian defined by  
\begin{equation}
{\cal H}_{\rm tot}\equiv \lim _{\lambda \rightarrow \infty}\left(\frac{{\cal H}_{\lambda}-E_{0}}{\lambda}\right)
\label{def},
\end{equation}
it was shown \cite{Poly2,S2,Kato} that 
\begin{equation}
{\cal H}_{\rm tot}={\cal H}_{\it t\mbox{-}J}+{\cal H}_{\rm ph}.\label{decouple}
\end{equation}
In eq. (\ref{decouple}), ${\cal H}_{\rm ph}$ is Hamiltonian of free phonons: 
\begin{equation}
{\cal H}_{\rm ph}=
\frac{\pi^2 J}{N^2}
\sum_{q=0}^{N-1}
q\left(N-q\right)\left(\hat b_{q}^{\dagger}\hat
b_{q}+\frac12\right),\label{elastic2}
\end{equation}
where $J=2\left(N/L\right)^2$
 and $\hat b_{q}^{\dagger}\,
 \left(\hat b_{q}\right)$ is the creation (annihilation ) operator of a phonon. 
On the other hand ${\cal H}_{\it t\mbox{-}J}$ describes the SU($K$,1) {\it t-J} model:
\begin{equation}
{\cal H}_{\it t\mbox{-}J}=-\frac{J}{2}\sum_{i<j}D^{-2}_{ij}\tilde P_{ij}, \mbox{  with  }D_{ij}\equiv \frac{N}{\pi}\sin \left(\frac{\pi\left(i-j\right)}{N}\right), \label{lattice}
\end{equation}
where the subscript $i$ and $j$ represent the lattice points. Fermions with spin $\sigma$ in the Sutherland model correspond to hard-core fermions with $\sigma$, and bosons to holes. In the SU(2,1) case, expression (\ref{lattice}) is equivalent to the following familiar form:
\begin{equation}
{\cal H}={\cal P}\sum_{i\ne j}\left[-t_{ij}\sum_{\sigma=\uparrow,\downarrow}c^{\dagger}_{i\sigma}c_{j\sigma}+\frac{J_{ij}}{2}\left(\mbox{\boldmath $S$}_{i}\cdot\mbox{\boldmath $S$}_{j}-\frac14 n_{i}n_{j}\right)\right]{\cal P}.
\end{equation}
Here $c_{i\sigma}$ is the annihilation operator of an electron with spin $\sigma$; $n_{i}$ and $\mbox{\boldmath $S$}_{i}$ are the number and spin operator, respectively. ${\cal P}\equiv \prod_{j}\left(1-c^{\dagger}_{j \uparrow}c_{j \uparrow}c^{\dagger}_{j \downarrow}c_{j \downarrow}\right)$ is the projection operator excluding the double occupation. The transfer energy $t_{ij}$ and exchange one $J_{ij}$ are given by $2t_{ij}=J_{ij}=J D^{-2}_{ij}$. From now on, we set $J=2$ for simplicity.

We note that thermodynamic quantities of ${\cal H}_{t\mbox{-}J}$ are obtained from those of ${\cal H}_{\tot}$ and ${\cal H}_{\rm ph}$ by eq. (\ref{decouple}). In the rest of this section, we construct thermodynamics for ${\cal H}_{t\mbox{-}J}$; first we consider thermodynamics for ${\cal H}_{\tot}$ and subsequently subtract the free phonon contribution. 

For the moment, we consider thermodynamics of ${\cal H}_{\rm tot}$. 
Using (\ref{Elambda}) and the relation (\ref{def}), we obtain the expression for the energy spectrum of ${\cal H}_{\tot}$\cite{S2,Kato}
\begin{equation}
E_{\rm tot}=\lim_{\lambda\rightarrow \infty}\left(\frac{E_{\lambda}-E_{0}}{\lambda}\right)=\frac{\pi^2}{N^2} \sum_{\kappa=\infty}^{\infty}\sum_{\kappa'=-\infty}^{\infty}\left|\kappa -\kappa '\right|\nu (\kappa )\nu (\kappa' ).\label{Efinite}
\end{equation}
Here we take the thermodynamic limit $N\rightarrow \infty$. In this limit, the energy is rewritten as
\begin{equation}
E_{\tot}=\frac{N}{8\pi}\int_{-\infty}^{\infty}\dk \intinf{\dk'}\left|k-k'\right|\nu(k)\nu(k'), \label{Etotk}
\end{equation}
where $\nu(k)$ is to be identified as $\nu(\kappa)$ with $\kappa=N k/(2\pi)$.
In eq. (\ref{Efinite}), we see that each microscopic eigenstate is uniquely characterized by the momentum distribution functions $\left\{\nu_{\sigma}(\kappa), \nu_{\B}(\kappa)\right\}$. Hence the entropy $s_{\rm tot}$ per site has the same form as that of the SU($K$,1) free particle system:
\begin{eqnarray}
s_{\rm tot}=\frac{1}{2\pi}\int_{-\infty}^{\infty}\dk\left\{\right.&-&\sum_{\sigma=1}^{K}\left[\left(1-\nu_{\sigma}\right)\ln \left(1-\nu_{\sigma}\right)+\nu_{\sigma}\ln\nu_{\sigma}\right]\nonumber\\
&+&
\left.
\left(1+\nu_{\B}\right)\ln \left(1+\nu_{\B}\right)
-\nu_{\B}\ln \nu_{\B }
\right\}.\label{entropy}
\end{eqnarray}
The density of fermions and bosons are given as
\begin{equation}
n_{\sigma}=\frac{1}{2\pi}\int_{-\infty}^{\infty}\dk \nu_{\sigma}\left(k\right)\label{number},\quad
n_{\B}=\frac{1}{2\pi}\int_{-\infty}^{\infty}\dk \nu_{\B}(k),\label{number2}
\end{equation}
respectively.

From eqs. (\ref{Etotk}), (\ref{entropy}) and (\ref{number2}), we obtain the expression for the thermodynamic potential: $\Omega_{\rm tot}(\left\{\nu\right\})/N=E_{\rm tot}
\left(\left\{\nu\right\}\right)/N
-Ts_{\rm tot}
\left(\left\{\nu\right\}\right)
-\sum_{\sigma}\mu_{\sigma}n_{\sigma}
\left(\left\{\nu\right\}\right)
-\mu_{\B}
n_{\B}
\left(\left\{\nu\right\}\right)$. 
The equilibrium conditions: $$
\delta \Omega(\left\{\nu_{\sigma},\nu_{\B}\right\})/\delta \nu_{\sigma}
=
\delta \Omega(\left\{\nu_{\sigma},\nu_{\B}\right\})/\delta \nu_{\B}=0$$
yields the equilibrium momentum distribution functions:
\begin{equation}
\nu_{\sigma}\left(\epsilon(k)-\mu_{\sigma}\right)=
\frac{1}{\exp\left[\left(\epsilon(k)-\mu_{\sigma}\right)/T\right]+1}, \label{nusigmadef}
\end{equation}
\begin{equation}
\nu_{\B}\left(\epsilon(k)-\mu_{\B}\right)=\frac{1}{\exp \left[\left(\epsilon(k)-\mu_{\B}\right)/T\right]-1}. \label{nu0def}
\end{equation}
Here we have introduced the one-particle energy $\epsilon (k)$ defined by
\begin{equation}
\epsilon (k)=\frac{2\pi}{N}\frac{\delta E_{\tot}(\left\{\nu _{\kappa}\right\})}{\delta \nu_{\kappa}}
=\frac{1}{2}\int_{-\infty}^{\infty}\dk'\left|k -k'\right|\nu (k').\label{epsilondef}
\end{equation}
Equation (\ref{epsilondef}) with (\ref{nusigmadef}) and (\ref{nu0def}) gives the functional equation for $\eps(k)$. Substituting the resultant expression for $\eps (k)$ into eqs. (\ref{Etotk})-(\ref{nu0def}), we can obtain thermodynamic quantities for ${\cal H}_{\rm tot}$. However it is tedious to solve eq. (\ref{epsilondef}) directly. In order to simplify the calculation, we introduce a new variable $p(k)$ defined as
\begin{equation}
p(k)\equiv\frac{\partial \epsilon (k)}{\partial k}
=\frac12\int_{-\infty}^{\infty}\dk' {\rm sgn}\left(k-k'\right)\nu (k'), \label{pk} \end{equation}
 which we call rapidity. In the following paragraphs, we will rewrite the thermodynamic quantities in terms of $p$. 

First we obtain an equation for $\eps$ as a function of $p$. Differentiation of eq. (\ref{pk}) with $k$ gives
\begin{equation}
\frac{\partial p(k)}{\partial k}=\nu (k)\label{dpdk}.
\end{equation}
By multiplying both sides of eq. (\ref{dpdk}) by $p(k)$ and integrating over $k$ from $k$ to $\infty$, we obtain the following relation:\
\begin{eqnarray}
\eps_{0}&=&\int ^{\infty}_{\epsilon}{\rm d}\epsilon' \nu (\epsilon ')\nonumber\\
&=&-T\ln \left(1-\exp\left[-\left(\epsilon - \mu _{\B}\right)/T\right]\right)+T\sum_{\sigma =1}^{K}\ln \left(1+\exp\left[-\left(\epsilon -\mu _{\sigma }\right)/T\right]\right), 
\label{pe}
\end{eqnarray}
where $\eps_{0}=\eps_{0}(p)\equiv(\pi^2 -p^2)/2$ and $\nu(\eps)$ represents $\nu\left(k\left(\eps\right)\right)$.
In the derivation of eq. (\ref{pe}), we have used the boundary condition: 
\begin{equation}
p(k=\pm \infty)=\pm \pi, \label{bc}
\end{equation}
which is obtained from eq. (\ref{pk}). 

For later convenience, we introduce 
\begin{equation}
X(p)=\exp \left[-\left(\eps -\mu_{\B}\right)/T\right]\label{Xdef}\end{equation}
and rewrite eq. (\ref{pe}) as
\begin{equation}
\eps_{0}\left(p\right)=-T\ln\left[1-X\left(p\right)\right]+T\sum_{\sigma=1}^{K}\ln\left[1+X\left(p\right)\exp\left(\zeta_{\sigma}/T\right)\right],\label{epsX} 
\end{equation}
with $\zeta_{\sigma}\equiv \mu_{\sigma}-\mu_{\B}$. $X(p)$ is the solution satisfying $X(p=\pi)=0$. Equation (\ref{epsX}) determines $X(p)$ for given $p$, $T$, $\zeta_{\sigma}$. In the rest of this section, we rewrite the expressions for the thermodynamic quantities in terms of $p$, $T$, and $\zeta_{\sigma}$. Here note that $\zeta_{\sigma}=\mu_{\sigma}-\mu_{\B}$ is the chemical potential for election with spin $\sigma$; introduction of an electron is equivalent to removing a boson and adding a fermion.   

Now we consider the expression for $E_{\tot}$. The constraint $n_{\B}+\sum_{\sigma=1}^{K}n_{\sigma}=1$ leads the expression for $\mu_{\B}$:
\begin{equation}
\mu_{\B}=T\ln X\left(p=0,T,\left\{\zeta_{\sigma}\right\}\right)+\int_{0}^{\pi}\ddp \frac{\pi -p}{\nu\left(p\right)}, \label{muB}
\end{equation} 
where $\nu(p)$ is the abbreviation of $\nu\left(k\left(p\right)\right)$. We will derive eq. (\ref{muB}) in Appendix A. 
From eqs. (\ref{Xdef}) and (\ref{muB}), we obtain the expression for the one-particle energy $\eps(p)$
\begin{equation}
\eps(p)=-T\ln\left[\frac{X(p)}{X(0)}\right]+\int_{0}^{\pi}\ddp\frac{\pi-p}{\nu(p)}. \label{epsexp}
\end{equation}
By using (\ref{dpdk}), (\ref{bc}), and (\ref{epsexp}), the internal energy (\ref{Etotk}) is written as
\begin{equation}
\frac{2\pi E_{\tot}}{N}=\int_{0}^{\pi}\ddp \eps(p)=-T\int_{0}^{\pi}\ddp \ln\left[\frac{X(p)}{X(0)}\right]+\pi\int_{0}^{\pi}\ddp\frac{\pi-p}{\nu(p)}.
\label{Etot1}
\end{equation}
Using the following mathematical relation:
\begin{equation}
\int_{0}^{\pi}\ddp \ln\left[\frac{X(p)}{X(0)}\right]=-\frac{1}{T}\int_{0}^{\pi}\ddp\frac{p(\pi-p)}{\nu(p)}, \label{MATH}
\end{equation} 
we find a simple form for $E_{\rm tot}$: 
\begin{equation}
\frac{2\pi E_{\rm tot}}{N}=\int_{0}^{\pi}\ddp \frac{\pi^2 - p^2}{\nu(p)}=\int_{-\pi}^{\pi}\ddp \frac{\eps_{0}(p)}{\nu(p)}. \label{Etot}
\end{equation}
The relation (\ref{MATH}) will be proven in Appendix B. 
Similarly we can represent density and entropy in simple forms. By using (\ref{dpdk}), the electron density $n_{\sigma}$ with $\sigma$ spin is written as
\begin{equation}
n_{\sigma}=\frac{1}{2\pi}\int_{-\pi}^{\pi}\ddp \frac{\nu_{\sigma}(p)}{\nu(p)}. \label{ne}
\end{equation}
The entropy (\ref{entropy}) per site is rewritten as
\begin{equation}
2\pi s_{\tot}=2\pi \left[E_{\tot}/N-\left(\sum_{\sigma=1}^{K}\zeta_{\sigma}n_{\sigma}\right)\right]/T-\int_{-\pi}^{\pi}\ddp \ln X(p)
\label{Stot},   
\end{equation}
which will be derived in Appendix C. 

So far we have constructed thermodynamics for ${\cal H}_{\rm tot}$. In order to obtain thermodynamic quantities for ${\cal H}_{t\mbox{-}J}$, we should subtract the free phonon contribution from the thermodynamic quantities for ${\cal H}_{\rm tot}$. The thermodynamic potential $\Omega_{\rm ph}$ for ${\cal H}_{\rm ph}$ (eq. (\ref{elastic2})) is given by
\begin{equation}
2\pi\Omega_{\rm ph}/N=T\int_{-\pi}^{\pi}\ddp \ln\left[1-\exp \left(-\eps_{0}(p)/T\right)\right]+\pi^3/3.\label{omegaph}
\end{equation}
The entropy per site $s_{\rm ph}$ and the internal energy $E_{\rm ph}$ are obtained from the expression (\ref{omegaph}). The internal energy for the {\it t-J} model is now given by
\begin{eqnarray}
\frac{2\pi E_{t\mbox{-}J}}{N}&=&\frac{2\pi(E_{\tot}-E_{\ph})}{N}\nonumber\\
&=&\int_{-\pi}^{\pi}\ddp \eps_{0}(p)\left(\frac{1}{\nu(p)}-\frac{1}{\exp [\eps_{0}(p)/T]-1}\right)-\frac{\pi^3}{3}.
\end{eqnarray}
The thermodynamic potential $\Omega_{t\mbox{-}J}$ is represented by
\begin{eqnarray}
\frac{2\pi \Omega_{\it t\mbox{-}J}}{N}&=&2\pi\left(\frac{E_{\tot}}{N}-T s_{\tot}-\sum_{\sigma=1}^{K}\zeta_{\sigma} n_{\sigma}-\frac{\Omega_{\ph}}{N}\right)\label{Omegatotph}\\
&=&T\int_{-\pi}^{\pi}\ddp
\ln \left(\frac{X(p)}{1-\exp\left[-\eps_{0}\left(p\right)/T\right]}\right)-\frac{\pi^3}{3}.
\label{OmegatJ}
\end{eqnarray}
The expression (\ref{OmegatJ}) is a main result in this section and will play crucial roles in section 5 and 6. It provides a practical procedure to calculate the thermodynamic quantities for the SU(2,1) case in section 5. It also gives a starting point of the calculation in section 6, where we derive the low temperature description for SU($K$,1) model. 

In summary of this section, the thermodynamic potential for the SU($K$,1) {\it t-J} model (\ref{lattice}) is given by (\ref{OmegatJ}). Here $X(p)$ is the solution of eq. (\ref{epsX}) satisfying the condition $X(p=\pi)=0$.
\section{Equivalence to Free Particles Obeying Fractional Exclusion Statistics in SU(2,1) Case}

In the previous section, the thermodynamic potential for the SU($K$,1) Sutherland model were obtained in a simple and closed form. However, it is hard to obtain the physical picture of the thermodynamic quantities from eq. (\ref{OmegatJ}). Hence it is desirable to have an alternative description. 
For the spin model with the $1/\sin ^2$ interaction, Haldane \cite{Haldane91} has already given thermodynamics in terms of free spinons obeying semionic statistics.
Here, we extend this type of description to the SU(2,1) {\it t-J}  model and construct thermodynamics in terms of free spinons and holons. 

In this section, we first introduce the statistics heuristically, and show that the spinons and holons with a particular exclusion statistics reproduces the thermodynamics obtained in the previous section.  Thus the picture of free spinons and holons turns out to be exact.
In this and following two sections, we represent the electron density by $n_{\rm e}\equiv n_{1}+n_{2}$, magnetization by $m\equiv n_{1}-n_{2}$, electron chemical potential by $\zeta\equiv \left(\zeta_{1}+\zeta_{2}\right)/2$, and magnetic field by $h\equiv (\zeta_{1}-\zeta_{2})/2$.

Let us consider a system consisting of three species of free particles: up and down spinons, and holons. We require that the energy including the magnetic energy and chemical potential is given by
\begin{equation}
2\pi \left(\frac{E}{N}-mh-\zeta n_{\rm e}+\zeta\right)=\frac{1}{2}\int_{-\pi}^{\pi}\ddp\left(\eps_{\up}\rho_{\up}+\eps_{\dn}\rho_{\dn}+\eps_{\rm h}\rho_{\rm h}\right),
\end{equation}
where $\eps_{\uparrow}=\eps_{0}(p)/2-h $ is the energy for up spinon and $\eps_{\downarrow}=\eps_{0}(p)/2+h $ for down spinon. We take $\eps_{0}=\eps_{0}(p)=(\pi^2 -p^2)/2$ to make correspondence with the SU(2,1) {\it t-J}  model.  
The energy of the holon is given by 
\begin{equation}
\eps_{\rm h}(p)=\zeta-\eps_{0}(p)/2\label{epsh}.
\end{equation}
The distribution functions for spinons and holons are given by $\rho_{\sigma}(p)$ and $\rho_{\rm h}(p)$, respectively.

We next introduce the distribution functions for dual particles (holes) by $\rho^{*}_{\al}$ for $\alpha = \up$, $\dn$, and h. 
Then the entropy is given by
\begin{equation}
2\pi s=\frac12 \sum_{\al=\up, \dn, {\rm h}}\int_{-\pi}^{\pi}\ddp\left[\left(\rho_{\al}+\rho^{*}_{\al}\right)\ln\left(\rho_{\al}+\rho^{*}_{\al}\right)-\rho_{\al}\ln \rho_{\al}-\rho^{*}_{\al}\ln\rho^{*}_{\al}\right]\label{threeent}.
\end{equation}
The exclusion statistics \cite{Haldanefrac,Wu} of each particle is given by the following relation:
\begin{equation}
\rho^{*}_{\al}=1-\sum_{\al'}g_{\al \al'}\rho_{\al'}\label{rhohrho},
\end{equation}
where the statistical matrix $g_{\al \al'}$ has been introduced.
Equations (\ref{threeent}), (\ref{rhohrho}) give the thermodynamics of free particles obeying the fractional exclusion statistics . 

By taking the functional derivative of the thermodynamic potential: $\Omega/N=E/N-mh-\zeta n_{\rm e}-Ts$ with respect to $\rho_{\al}$, we obtain the following constitutive equations for the three components:
\begin{equation}
\eps_{\al}/T=\ln\left(1+w_{\al}\right)-\sum_{\al'}g_{\al' \al}\ln\left(1+w^{-1}_{\al'}\right), \label{therm}
\end{equation}
where $w_{\al}\equiv\rho^{*}_{\al}/\rho_{\al}$.
From eqs. (\ref{rhohrho}) and (\ref{therm}), we can obtain the thermal distribution function for each $\al$. In terms of $w_{\al}$, the thermodynamic potential is written as
\begin{equation}
2\pi\left(\frac{\Omega}{N} +\zeta\right)=-\frac{T}{2}\int_{-\pi}^{\pi}\ddp\sum_{\al}\ln\left(1+w^{-1}_{\al}\right).\label{OmegaFES}
\end{equation}

If we set the following values for the statistical matrix: 
\begin{equation}
\{g_{\al \al'}\}=\left(
\begin{array}{ccc}
1/2&1/2&-1/2\\
1/2&1/2&-1/2\\
1/2&1/2&1/2
\end{array}
\right), \label{matrix}
\end{equation}
equations (\ref{therm}) and (\ref{OmegaFES}) yield the thermodynamic potential (\ref{OmegatJ}) with $K=2$. Here we show it by a direct calculation.

By using eq. (\ref{therm}) for $\alpha =\uparrow $, we rewrite (\ref{OmegaFES}) as
\begin{equation}
2\pi \Omega/N=-2\pi \zeta+\eps_{\up}-\ln\left(1+w_{\up}\right).\label{direct1}
\end{equation} 
With a new quantity $\tilde X(p)=\tilde X(p,T,\zeta,h)$ defined by the relation:
\begin{equation}
1+\wup= \frac{\exp\left(\eps_{0}/T\right)-1}{\tilde X(p)\exp\left[\left(h+\zeta\right)/T\right]}, 
\label{Xtilde}
\end{equation}
expression (\ref{direct1}) is rewritten as
\begin{equation}
\frac{2\pi \Omega}{N}=T\int_{-\pi}^{\pi}\ddp
\ln \left(\frac{\tilde X(p)}{1-\exp\left[-\eps_{0}\left(p\right)/T\right]}\right)-\frac{\pi^3}{3}.
\label{direct2}
\end{equation}
Expression (\ref{direct2}) has the same form as (\ref{OmegatJ}), if $X(p)=\tilde X(p)$. In the following we show that $X(p)=\tilde X(p)$.
 
The constitutive equations (\ref{therm}) with (\ref{matrix}) turn into a quadratic equation for $w_{\al}$; for $\wup$, we obtain the following equation:
\begin{equation}
w_{\up}+\left[1-e^{-2h/T}-e^{\left(2\eps_{\up}+h-\zeta\right)/T}\right]w_{\up}-e^{2\eps_{\up}/T}\left[1+e^{\left(h-\zeta\right)/T}\right]=0. 
\label{quadratic}
\end{equation}  
From eqs. (\ref{Xtilde}) and (\ref{quadratic}), we can derive the following relation: 
\begin{equation}
\eps_{0}=-T\ln\left[1-\tilde X\left(p\right)\right]+
T\ln\left(1+\tilde X\left(p\right)\exp\left[\left(\zeta +h\right)/T\right]\right)+T\ln\left(1+\tilde X\left(p\right)\exp\left[\left(\zeta -h\right)/T\right]\right).\label{direct3}
\end{equation}
From (\ref{Xtilde}), we can see easily that $\tilde X(p=\pi)=0$. Thus $\tilde X(p)$ is nothing but $X(p)$ and hence thermodynamic potential in eq. (\ref{direct2}) is the same as that in (\ref{OmegatJ}) for SU(2,1) case.  

Now we find that the {\it t-J}  model is equivalent to the system of free particles obeying the fractional exclusion statistics with the statistical parameters (\ref{matrix}). This is a main result in this section. 
In the rest of this section, we discuss the meaning of the expression (\ref{matrix}). 

Consider an $N$ site system with $N_{\up}$ up spinons, $N_{\dn}$ down spinons, and $N_{\rm h}$ holons.  To be consistent with the statistical matrix (\ref{matrix}), the numbers of available one-particle states for $\up$, $\dn$, and $\rm h$ should be given by 
\begin{subequations}
\begin{equation}
D_{\up}=D_{\dn}=1+\frac12 (N-N_{\up}-N_{\dn}+N_{\rm h})\label{dup}
\end{equation}
\begin{equation}
D_{\rm h}=1+\frac12 (N-N_{\up}-N_{\dn}-N_{\rm h})\label{dh}.
\end{equation}
\end{subequations}
By the relation $\partial D_{\al}/\partial N_{\beta}=-g_{\al \beta}$, \cite{Haldanefrac} we can easily reproduce eq. (\ref{matrix}).
In the absence of holons {\it i.e.} $N_{\rm h}=0$, the expression (\ref{dup}) reproduces the fractional exclusion statistics of spinons \cite{Haldanefrac}, as it should.  
We now show that spinons and holons span the full Hilbert space of hard core electrons where no double occupation of a site is allowed. 
With $N_{\up}$, $N_{\dn}$, $N_{\rm h}$ fixed, the number of many spinon-holon states is given by
\begin{equation}
W\left(N,N_{\up},N_{\dn},N_{\rm h}\right)=\frac12
\left[1+\left(-1\right)^{N-N_{\up}-N_{\dn}-N_{\rm h}}\right]
\frac{\left(D_{\up}+N_{\up}-1\right)!}{\left(D_{\up}-1\right)!N_{\up}!}
\frac{\left(D_{\dn}+N_{\dn}-1\right)!}{\left(D_{\dn}-1\right)!N_{\dn}!}
\frac{\left(D_{\rm h}+N_{\rm h}-1\right)!}{\left(D_{\rm h}-1\right)!N_{\rm h}!}.\label{state}
\end{equation}
where the factor $\left[1+\left(-1\right)^{N-N_{\up}-N_{\dn}-N_{\rm h}}\right]/2$ picks out even numbers of spinon-holon excitations for even $N$, and odd numbers of excited particles for odd $N$.  We remind that the pair creation is an intrinsic property of kink excitations.  
With eq. (\ref{state}), the dimension of the full Hilbert space is given by
\begin{equation}
\sum_{0\le N_{\up}+N_{\dn}+N_{\rm h}\le N}W\left(N_{\up},N_{\dn},N_{\rm h}\right), 
\end{equation} 
which amounts to $3^{N}$.  Therefore the spinons and the holons exhaust the Hilbert space for the {\it t-J} model.

The statistical matrix given by eq. (\ref{matrix}) is not the unique choice in spanning the complete Hilbert space. 
For example an alternative choice:
\begin{equation}
\{g_{\al \al'}\}=\left(
\begin{array}{ccc}
1/2&1/2&1/2\\
1/2&1/2&1/2\\
0&0&1
\end{array}
\right)\label{matrix2}
\end{equation}
can also span the full Hilbert space.  In this case, the holon behaves as a fermion and need not be excited in pairs.
Obviously the counting of the dimension of the Hilbert space does not depend on parameters in a Hamiltonian.
The important point is that particles obeying the alternative statistics defined by eq. (\ref{matrix2}) are not free in the supersymmetric {\it t-J} model.
 Only with the choice eq. (\ref{matrix}) the excitations behave as free particles in the model.  
\section{g-on Description for Low Temperature Properties of SU(2,1) Model. }
In the previous section, we constructed the thermodynamics of the SU(2,1) model in terms of spinons and holons. In the scheme, spinons and holons are free energetically but interact with each other via the statistical interaction. At low temperatures, however, the statistical interaction between them plays no crucial roles and, as a result, both spinons and holons obey the g-on statistics. The spin-charge separation at zero temperature has been observed in ref.(\citen{KY}). In this section, we show that the spin-charge separation still holds up to a certain characteristic temperature. 

Figure (1$\cdot$a) shows distribution functions of spinons and holons at $h=0$. Here $p_{\rm h} = \left(\pi^2 -4 \zeta\right)^{1/2}$. Both spinons and holons exist only in the region $0\le \left| p \right|\le \pi$. However we extend the abscissa to the outside of the region, in order to emphasize that the spinon distribution function turns into the fermionic distribution function. In Fig. (1$\cdot$a), we can see that spinons and holons distribute in separate regions in the $p$ space. From eqs. (\ref{threeent}) and (\ref{rhohrho}), the statistical interaction between spinons and holons exist only for the same $p$. Hence spinons and holons decouple from each other in thermodynamics at low temperatures. This simple fact indicates the spin-charge separation in this model. At low temperatures, holons and spinons can be regarded as g-ons \cite{Wu,Nayak,Murphy}, which interact with each other statistically only within the same species. 

Figure (1$\cdot$b) shows the distribution function of up spinons and holons in the presence of magnetic field ($h >0$). Here $p_{\uparrow}$ is given by $\left(\pi^2 -4 h\right)^{1/2}$. In this case, few down-spinons are excited at low temperatures, since down spinons have a gap in the spectrum. In Fig. (1$\cdot$b), we observe that the spin-charge separation occurs also in the presence of a magnetic field. 
From Fig. (1$\cdot$a) (1$\cdot$b), we find that the g-on description depends on the value of $h$. Hence we consider the two cases $h=0$ and $h \ne0$ separately in the following subsections. 
\subsection{Case of $h=0$}
In this case, we define the low temperature region such that the following relation is satisfied:
\begin{equation}
\exp\left(-\zeta/T\right)\ll 1.\label{Deltac} 
\end{equation} 
In the absence of magnetic field, we can set $\eps_{\up}=\eps_{\dn}\equiv \eps_{\rm s}$, $\rho_{\up}=\rho_{\dn}\equiv \rho_{\rm s}$, and $w_{\up}=w_{\dn}\equiv w_{\rm s}$. Eliminating $w_{\rm h}$ in eq. (\ref{therm}), we obtain 
\begin{equation}
w_{\rm s}^2\exp(-2\eps_{\rm s}/T)-1=\left(1+w_{\rm s}\right)\exp(-\zeta /T). \label{wseqn}
\end{equation}
Now we define the rapidity region $\Pis$ as the region where the following condition is satisfied:
\begin{equation}
\ws \exp(-\zeta/T)\le \eta,\label{Piscond}
\end{equation}
where $\eta$ is a constant which is much smaller than unity but much larger than $\exp(-\zeta/T)$. In $\Pis$, the solution of eq. (\ref{Deltac}) is given by
\begin{equation}
\ws \simeq \exp(\eps_{\rm s}/T)\quad\mbox{for }p\in\Pis.\label{wsPis}
\end{equation}
With the use of eq. (\ref{wsPis}), we can rewrite the inequality (\ref{Piscond}) as
\begin{equation}
\exp[(\eps_{\rm s}-\zeta)/T]\le \eta\quad \mbox{for }p\in \Pis.
\end{equation}
By definition, the following inequalities are satisfied in the region ${\overline \Pis}\equiv \left[-\pi,\pi\right]-\Pis$ complementary to $\Pis$:
\begin{equation}
\ws >\eta \exp\left(\zeta/T\right), 
\end{equation}
\begin{equation}
\exp(-\eps_{\rm s}/T)<\eta^{-1}\exp\left(-\zeta/T\right)\quad\mbox.
\end{equation}
The contribution from the region ${\overline \Pis}$ to the thermodynamic potential is negligible because
\begin{equation}
0<\int_{\overline \Pis}\ddp \ln\left(1+\ws^{-1}\right)<\int_{\overline \Pis}\ddp\ln\left[1+\eta^{-1}\exp\left(-\zeta/T\right)\right]\ll 1.
\end{equation}
Hence the spinon part of the thermodynamic potential becomes
\begin{equation}
2\pi\left(\Omega_{\rm s}/N\right)\equiv
-T\int_{-\pi}^{\pi}\ddp \ln\left(1+w^{-1}_{\rm s}\right)\simeq -T\int_{\Pis}\ddp\ln\left[1+\exp\left(-\eps_{\rm s}/T\right)\right]. \label{Omegas} 
\end{equation}
Furthermore, we can extend the range of integration in eq. (\ref{Omegas}) to $[-\pi,\pi]$ because
\begin{equation}
\int_{\overline \Pis}\ddp\ln\left[1+\exp\left(-\eps_{\rm s}/T\right)\right]<\int_{\overline \Pis}\ddp\ln\left[1+\eta^{-1}\exp\left(-\zeta/T\right)\right]\ll 1.
\end{equation}
As a result, we obtain the low temperature expression for the spinon part in the thermodynamic potential as
\begin{equation}
2\pi\left(\Omega_{\rm s}/N\right)\simeq -T
\int_{-\pi}^{\pi}\ddp \ln\left[1+\exp\left(-\eps_{\rm s}/T\right)\right]\label{Omegas2}. 
\end{equation}
From the expression (\ref{Omegas2}), we find that the spinon part behaves as a free fermion one with energy $\eps_{\rm s}$ in the low temperature region.   This is because spinons are always excited in pairs without participation of holons.

Similarly, we obtain the low temperature description for the holon part in thermodynamics.  We eliminate $\ws$ in eq. (\ref{therm}) and obtain
\begin{equation}
\exp\left(\frac{2\eps_{\rm h}}{T}\right)=\frac{\wh\left(1+\wh\right)}{\left[1-\wh \exp\left(-\zeta/T\right)\right]^2}\label{wceqn}.
\end{equation}
We define the rapidity region $\Pic$ as the region where $\wh <\eta \exp\left(\zeta/T\right)$ is satisfied. As in the case of spinons, main contribution to the holon part in the thermodynamic potential comes from the region $\Pic$. In $\Pic$, we obtain 
\begin{equation}
\exp\left(\eps_{\rm h}/T\right)\simeq \wh^{1/2}\left(1+\wh\right)^{1/2}\quad\mbox{for }p\in\Pic.\label{whsolution}
\end{equation}
Here we introduce $\wc=\wc(p)$ as the positive solution of the following equation:
\begin{equation}
\exp\left(\eps_{\rm h}/T\right)= \wc^{1/2}\left(1+\wc\right)^{1/2}\quad\mbox{for }p\in \left[-\pi,\pi\right].\label{wcsolution}
\end{equation}
In terms of $\wc$, the holon part in the thermodynamic potential is rewritten as\begin{equation}
2\pi\left(\frac{\Omega_{\rm c}}{N}+\zeta\right)\equiv -\frac{T}{2}\int_{-\pi}^{\pi}\ddp\ln\left(1+\wh^{-1}\right)\simeq -\frac{T}{2}\int_{-\pi}^{\pi}\ddp\ln\left(1+\wc^{-1}\right). \label{Omegawc}
\end{equation}
From eqs. (\ref{whsolution}) and (\ref{Omegawc}), we find that the holon behaves as a semion with the energy $\eps_{\rm h}$ in the low temperature region. Since the spinon part (\ref{Omegas2}) does not contain $\zeta$, the charge susceptibility $\chi_{\rm c}$ is determined solely by the holon distribution function at low temperatures:
\begin{equation}
\chi_{\rm c}\equiv \left.\frac{\partial n_{\rm e}}{\partial \zeta}\right|_{h=0}
= -\frac{1}{4\pi}\int_{-\pi}^{\pi}\ddp \frac{\partial \rho_{\rm h}}{\partial \zeta}\simeq -\frac{1}{4\pi}\int_{-\pi}^{\pi}\ddp \frac{\partial }{\partial \zeta}\left(w_{\rm c}+1/2\right)^{-1}.\label{chicc}
\end{equation}
Equation (\ref{chicc}) manifests the spin-charge separation at low temperatures.
\subsection{Case of $h\ne 0$}

In this subsection, we discuss the case where $h>0$ and $\exp\left(\zeta/T\right) \gg \exp\left(h/T\right) \gg 1$.  The holon part $\wh$ takes the same expression as in the previous subsection, while each spinon shows a  behavior different from each other.  Namely, the down spinon does not take part in the low temperature thermodynamics because its energy spectrum $\eps_{\dn}(p)$ is gapful.  The up spinon, on the other hand, turns into a free semion with energy $\eps_{\up}(p)$.  
Eliminating $\wh$ and $\wup$ in eq. (\ref{therm}), we obtain the following equation for $\wdn$:
\begin{equation}
\wdn^2+\left[1-e^{2h/T}-e^{\left(2\eps_{\dn}-h-\zeta\right)/T}\right]\wdn-e^{2\eps_{\dn}/T}\left[1+e^{-\left(h+\zeta\right)/T}\right]=0\label{wdneqn}.
\end{equation}
Since $\wdn$ is the positive solution of eq. (\ref{wdneqn}), we obtain the following relation:
\begin{equation}
\wdn >\left[e^{\left(2\eps_{\dn}-h-\zeta\right)/T}+e^{2h/T}-1\right]/2>\left(e^{2h/T}-1\right)/2\gg 1.\label{wdnrelation}
\end{equation}
From the relation (\ref{wdnrelation}), the down spinon part in the thermodynamic potential is estimated as
\begin{equation}
0<\int_{-\pi}^{\pi}\ddp\ln\left(1+w^{-1}_{\dn}\right)\simeq {\cal O}(e^{-2h/T})\ll1.
\end{equation}
Hence it is negligible in the low temperature region. 
The equation for $\wup$ obtained from eq. (\ref{therm}) is 
\begin{equation}
\exp\left(\frac{2\eps_{\up}}{T}\right)=\frac{\wup\left[\wup +1-\exp\left(-2h/T\right)\right]}{1+\left(\wup +1\right) \exp\left[\left(h-\zeta\right)/T\right]}. \label{wuprelation}
\end{equation}
As in the previous subsection, we define the rapidity region $\Piup$ as the region where  
the following inequalities are satisfied:
\begin{equation}
\wup \exp[\left(h-\zeta\right)/T]<\eta, 
\label{lowupdef}
\end{equation}
\begin{equation}
\wup \exp\left(-2h/T\right)<\eta. 
\end{equation}
Here we have introduced a constant $\eta$ which satisfies the inequalities $\exp[-\left(\zeta-h\right)/T]\ll \eta \ll 1$.
In $\Piup$, the equation (\ref{wuprelation}) becomes
\begin{equation}
\exp\left(\eps_{\up}/T\right)\simeq\wup^{1/2}\left(1+\wup\right)^{1/2}\quad\mbox{for }p\in \Piup.\end{equation}
The quantity $\ws$ is the solution of 
\begin{equation}
\exp\left(\eps_{\up}/T\right)=\ws^{1/2}\left(1+\ws\right)^{1/2}\quad\mbox{for }p\in [\pi,\pi].\label{wsdef}\end{equation}
Then the up spinon contribution to the thermodynamics can be written as
\begin{equation}
\frac{2\pi\Omega_{\up}}{N}\equiv-\frac{T}{2}\int_{-\pi}^{\pi}\ddp\ln\left(1+\wup^{-1}\right)\simeq -\frac{T}{2}\int_{-\pi}^{\pi}\ddp \ln\left(1+\ws^{-1}\right).\label{Omegaup2}
\end{equation}
From eqs. (\ref{wsdef}) and (\ref{Omegaup2}), we find that the up spinon behaves as a semion with the energy $\eps_{\rm s}(p)$ in the low temperature region defined by eq. (\ref{lowupdef}). 
\section{Numerical Results}
In this section, we present explicit results for the spin susceptibility $\chi_{\rm s}$, charge susceptibility $\chi_{\rm c}$ and the specific heat $C$.  First we obtain the integral representation for these quantities, by using the results in sections $2$ and $3$. Then we perform the integral numerically. We interpret the explicit results in terms of the g-on picture presented in the last section.  
We will describe temperature dependence of the three quantities, with the electron concentration $n_{\rm e}$ and the magnetization $m$ being fixed. Hence $\zeta$ and $h$ should be regarded as functions of $n_{\rm e}$ and $m$.

Figure \ref{chis} shows the spin susceptibility: 
\begin{equation}
\chi_{\rm s} = \frac14 \left.\frac{\partial m}{\partial h}\right|_{h=0}
\end{equation}
as a function of $T$ with $n_{\rm e}=$1.0, 0.9, 0.8, 0.6, and 0.4 for the electron density.  Peak structure in $\chi_s$, which is a characteristic of antiferromagnetic correlation, can be seen for all concentrations.  
As the density $n_{\rm e}$ decreases from unity, the spin susceptibility $\chi_s$ decreases in the high $T$ region.  This simply reflects the decrease of the number of local moments.
As $T \rightarrow 0$, the spin susceptibility approaches $\left(2 \pi^2\right)^{-1}$ irrespective of the electron concentration. Even at low temperatures, $\chi_{\rm s}$'s for various $n_{\rm e}$ collapse onto a single curve. Such behavior has not been observed in other models such as the Hubbard model \cite{Usuki} and the {\it t-J}  model with the nearest neighbor interaction \cite{Ogata}. Thus we call this behavior ^^ ^^ the strong spin-charge separation". \cite{KuramotoKato} As temperature increases, thermally excited holon begins to contribute to the spin susceptibility and then it begins to depend on the electron concentration.  The distribution functions for spinons and holons begin to overlap at a temperature where the tails of each distribution function have appreciable values at the ^^ ^^ Fermi surface" of each other.   We define the crossover temperature $T_{\times}$ by 
\begin{equation}
\rho_{\rm c}(p=p_{\rm s}=\pi)=0.01.\label{cross}
\end{equation}
We can see from Fig. \ref{chis} that $T_{\times}$ serves as a good scale for the crossover phenomena. 

Figure \ref{chic} shows temperature dependence of the charge susceptibility:  
\begin{equation}
\chi_{\rm c} =\left.\frac{\partial n_{\rm e}(\zeta,h=0,T)}{\partial \zeta}\right|_{T}
\end{equation}
for the various electron concentrations.
The peak structure in $\chi_{\rm c}$ can be seen in the low temperature regime for all concentrations $n_{\rm e}=0.2 \sim 0.9$. As $T \rightarrow 0$, the charge susceptibility approaches $2/[\pi^2 \left(1-n_{\rm e}\right)]$. $\chi_{\rm c}$ is enhanced anomalously as the system approaches the half-filling, while $\chi_{\rm c}$ vanishes just at the half-filling where the charge gap is present. This singularity in $\chi_{\rm c}$ is a typical behavior near the Mott transition and has been observed also in the Hubbard model \cite{Usuki} and the nearest neighbor {\it t-J}  model \cite{Ogata}. Above $T \sim 1.2$, we can observe the particle-hole symmetry $\chi_{\rm c} (T,n_{\rm e}=0.2)\sim \chi_{\rm c}(T,n_{\rm e}=0.8)$ and $\chi_{\rm c} (T,n_{\rm e}=0.4)\sim \chi_{\rm c}(T,n_{\rm e}=0.6)$.  This symmetry is expected for localized electrons with $\chi_{\rm c} = n_{\rm e} (1-n_{\rm e})/T$.  
Hence we identify $T \ge 1.2$ as the high temperature region. 

Now we study to what extent the g-on picture accounts for the low temperature thermodynamics. The holon is responsible for the charge excitation in the g-on description.  In Fig. \ref{chicne08} the solid curve represents the exact result on $\chi_{\rm c}$, which is the same as that in Fig. \ref{chic} at $n_{\rm e}=0.8$.  On the other hand, the dotted curve in Fig. \ref{chicne08} represents the charge susceptibility of free holons which is given by Eq. (\ref{chicc}). The crossover temperature $T_{\times}$ for $n_{\rm e}=0.8$ is denoted by $\times$. Below $T_{\times}$, these two curves are in good agreement with each other. Even above $T_{\times}$, the deviation is small and the free holon picture works well up to $T \sim 1.5$. Peak structure around $T \sim 0.06$ comes from the van Hove singularity of the free holon band (\ref{epsh}). 

Figure \ref{heat}(a) shows temperature dependence of the specific heat:  
\begin{equation}
C_{\it t\mbox{-}J } 
= \left.\frac{1}{N}\frac{\partial E_{\it t\mbox{-}J}}{\partial T}\right|_{n_{\rm e}},  
\end{equation}
for electron concentrations $n_{\rm e}=$1.0, 0.9, 0.8, 0.6, and 0.4. For all concentrations, peak structure can be seen around $T=0.7$. Near the half-filling, specific heat changes rapidly with temperature in the low temperature regime $T\le 0.1$. Figure \ref{heat}(b) blows up the specific heat in the low temperature region ($T =0 \sim 0.5$). The enhancement of $\gamma=C/T$ is evident for $n_{\rm e}=0.9$ and 0.8. 

In Fig. \ref{heatne08}, we compare the exact specific heat for the {\it t-J} model (solid curve) with the free holon contribution (dashed-dotted one), and the free spinon one (dotted one).  The sum of the specific heat of free holons and free spinons is also shown.  The deviation from the exact one indicates the contribution from the statistical interaction between spinons and holons.
The crossover temperature $T_{\times}$ defined in (\ref{cross}) is denoted by $\times$.
Up to $T \sim 0.3$, the additive contribution from holons and spinons accounts well for the exact result.  This confirms the spin-charge separation. We can attribute the rapid change below $T\sim 0.05$ to the free holon, which has the small rapidity $p_{\rm c}=\pi\left(1-n_{\rm e}\right)\sim 0.63$ at $n_{\rm e}=0.8$. 

\section{Low Temperature Description for SU$(K,1)$ Model}
Here we discuss the low temperature properties of the SU$(K,1)$ long-range {\it t-J}  model. The method in section 3 is not valid for the SU($K$,$1$) $(K\ge 3)$ model. Then we follow another route; first, we obtain the low temperature expression for the thermodynamics of ${\cal H}_{\rm tot}$, and then extract the contribution from the multicomponent {\it t-J}  model. 
We show that the internal energy $E_{\rm tot}$ can be written in the form of the sum of the g-on energy, in the low temperature regime defined as the region $ \exp(\zeta_{\sigma}/T)\gg 1$ for all $\sigma$. As in section 4, the low temperature description depends on the distribution of chemical potentials for each species. In subsection 6.1, we consider the polarized case with $\zeta_{1}>\zeta_{2}>\cdots >\zeta_{K}$. In 6.2, we consider the unpolarized case with $\zeta_{1}=\zeta_{2}=\cdots =\zeta_{K}\equiv \zeta$. 
\subsection{Polarized case}
In this case, the low temperature region is defined by the following conditions:
\begin{equation}
\exp(\zeta_{\sigma}/T)\gg 1\quad\mbox{for }1\le \sigma \le K
\end{equation}
and
\begin{equation}
\exp\left[\left(\zeta_{\sigma}-\zeta_{\sigma +1}\right)/T\right]\gg 1\quad \mbox{for } 1\le \sigma \le K-1. 
\end{equation}
The aim of this subsection is to derive eq. (\ref{Omegaunpo}). 
First we consider the rapidity dependence of the distribution function $\nu_{\B}(p)$ and $\nu_{\sigma}(p)$ at zero temperature. The distribution functions are given by
\begin{equation}
\nu_{\B}(p)=\left\{
\begin{array}{cc}
\infty&\quad\mbox{for }\left|p\right|<p_{\rm c}\equiv\left(\pi^2 -2\sum_{\sigma=1}^{K}\zeta_{\sigma}\right)^{1/2}\\
0&\quad\mbox{for }p_{\rm c}<\left|p\right|, 
\end{array}\right.
\label{nu0zero}
\end{equation}
\begin{equation}
\nu_{\sigma}(p)=\theta\left(p_{\sigma}-\left|p\right|\right),\quad\mbox{with }p_{\sigma}\equiv\left[\pi^2 -2\sum_{\alpha=1}^{\sigma-1}\zeta_{\alpha}+2\left(\sigma-1\right)\zeta_{\sigma}\right]^{1/2}. \label{nusigma} 
\end{equation}
Figure \ref{nuprofile} shows the rapidity dependence of $\nu(p)^{-1}$, which is a part of the integrand in eq. (\ref{Etot}), for the case of $K=2$. Near $p=p_{\al}$ $(\al={\rm c},1\sim K)$, $\nu_{\sigma}(\sigma\ne \al)$ can be regarded as a constant and hence the rapidity dependence of $\nu^{-1}$ is determined mainly by $\nu_{\al}$. Hence we can replace $\nu^{-1}$ by
\begin{eqnarray}
\nu^{-1}&\simeq&\nu^{-1}_{1}-\sum_{\sigma=2}^{K}\left(\frac{1}{\sigma-1}-\frac{1}{\sigma-1+\nu_{\sigma}}\right)-\left(\frac{1}{K}-\frac{1}{K+\nu_{\B}}\right)\nonumber\\
&=&1+\wFl-\sum_{\sigma=2}^{K}\frac{G^{2}_{\sigma}}{\wFs +g_{\sigma}}-\frac{G_{\rm c}^{2}}{\wB+g_{\rm c}},\label{nuinvapp} 
\end{eqnarray}
where $\wFs=\nu^{-1}_{\sigma}-1$, $\wB=\nu^{-1}_{\B}$, $g_{\sigma}=\sigma/\left(\sigma-1\right)$, $G_{\sigma}=\left(\sigma-1\right)^{-1}$ for $2 \le \sigma \le K$, and $g_{\rm c}= G_{\rm c} =K^{-1}$. 
The quantities $G_{\sigma}$ and $G_{\rm c}$ give the scaling of the momentum in the g-on description.  In the particular case of $K=2$, $G_{\rm c}=1/2$ means that the ^^ ^^ Brillouin zone" of the holon becomes half of the original size.

Let us relate the expression (\ref{nuinvapp}) with the distribution function of g-ons. First we rewrite eq. (\ref{epsX}) in terms of $\wFs$ and $\wB$ as
\begin{equation}
\exp \left(\eps_{0}/T\right)=\left(1+\wB^{-1}\right)\prod_{\sigma=1}^{K}\left(1+\wFs^{-1}\right). \label{wFswB}
\end{equation}
We eliminate $\wFs$ in eq. (\ref{wFswB}) with the use of the relation $\wFs=\left(1+\wB\right)\exp\left(-\zeta_{\sigma}/T\right)$. The resultant expression is 
\begin{equation}
\exp(\eps_{\rm c}/T)=\left(1+\wB\right)\left(1+\wB^{-1}\right)^{-1/K}\prod_{\sigma}\left[1+\left(1+\wB\right)\exp\left(-\zeta_{\sigma}/T\right)\right]^{-1/K},\label{wFswB2}
\end{equation}
with $\eps_{\rm c}=\left(\sum_{\sigma}\zeta_{\sigma}-\eps_{0}\right)/K$. 
Here we introduce the rapidity region $\Pic$ where $\wB \exp\left(-\zeta_{K}/T\right)<\eta \ll 1$ with $\eta$ a constant. In $\Pic$, equation (\ref{wFswB2}) reduces to
\begin{equation}
\exp\left(\eps_{\rm c}/T\right)\simeq \left(1+\wB\right)\left(1+\wB^{-1}\right)^{-1/K}=\wB^{1/K}\left(1+\wB\right)^{1-1/K}\quad \mbox{for }p\in \Pic. \label{wBeqn}
\end{equation}
Then we introduce the g-on quantities $\rhoc$ and $\wc$ as
\begin{equation}
\rhoc\equiv \left(\wc +g_{\rm c}\right)^{-1},
\end{equation}
\begin{equation}
\exp\left(\eps_{\rm c}/T\right)\equiv \left(1+\wc\right)^{1/K}\left(1+\wc^{-1}\right)^{1-1/K}\quad \mbox{for }p\in \left[-\pi, \pi\right]. 
\end{equation}
Apparently, 
\begin{equation}
\left(\wB +g_{\rm c}\right)^{-1}\simeq \rho_{\rm c}\label{wBrhoc}
\end{equation}
in $\Pic$. Since both sides of (\ref{wBrhoc}) are much smaller than unity outside $\Pic$, the relation (\ref{wBrhoc}) is valid for $p\in\left[-\pi, \pi\right]$. Hence we obtain the following expression for the boson part of the internal energy at low temperatures:
\begin{equation}
\int_{-\pi}^{\pi}\ddp\frac{\eps_{0}}{\wB +g_{\rm c}}\simeq \int_{-\pi}^{\pi}\ddp \eps_{0}\rhoc. \label{Energyc}
\end{equation}

Next we consider the term $\left(\wFs +g_{\sigma}\right)^{-1}$ in eq. (\ref{nuinvapp}). With the use of the relations: $\wB=\wFs \exp\left(\zeta_{\sigma}/T\right)-1$ and $\wFsp=\wFs \exp[\left(\zeta_{\sigma}-\zeta_{\sigma'}\right)/T]$, we rewrite eq. (\ref{wFswB}) as
\begin{equation}
\exp\left(\eps_{0}/T\right)=\left(1+\wFs^{-1}\right)\left[1-\wFs^{-1}\exp\left(-\zeta_{\sigma}/T\right)\right]^{-1}\prod_{\sigma'(\ne \sigma)}\left\{1+\wFs^{-1}\exp\left[\left(\zeta_{\sigma'}-\zeta_{\sigma}\right)/T\right]\right\}.\label{wFseqn}
\end{equation}
Here we define the rapidity region $\Pisigma$ as the region where the following relations are satisfied:
\begin{equation}
1 \ll  \wFs \exp(\zeta_{\sigma}/T)\quad\mbox{for }1 \le \sigma \le K,
\end{equation}
\begin{equation}
\exp\left[\left(\zeta_{\sigma'}-\zeta_{\sigma}\right)/T\right]\ll \wFs \ll \exp\left[\left(\zeta_{\sigma}-\zeta_{\sigma'}\right)/T\right]\quad\mbox{for }1 \le \sigma < \sigma' \le K. 
\end{equation}
In the region $\Pisigma$, equation (\ref{wFseqn}) reduces to
\begin{eqnarray}
\exp\left(\eps_{\sigma}/T\right)&\simeq& \wFs^{g_{\sigma}}\left(1+\wFs\right)^{1-g_{\sigma}}\quad\mbox{for }2\le \sigma \le K \\
\exp(\eps_{0}/T)&\simeq& 1+\wFl^{-1}
\end{eqnarray}
with $\eps_{\sigma}\equiv -\zeta_{\sigma}+G_{\sigma}\left(\sum_{\sigma'=1}^{\sigma-1}\zeta_{\sigma'}-\eps_{0}\right)$. 
Further we define the rapidity distribution function for g-ons as $\rho_{\sigma}=\left(\wsigma+g_{\sigma}\right)^{-1}$ for $2\le \sigma \le K$, where $\wsigma$ is given as the solution of\begin{equation}
\exp\left(\eps_{\sigma}/T\right)=\wsigma^{g_{\sigma}}\left(1+\wsigma\right)^{1-g_{\sigma}}\quad\mbox{for }p\in\left[-\pi,\pi\right]. 
\end{equation}
As in the boson case (eq. (\ref{Energyc})), the relation
\begin{equation}
\int \ddp\frac{\eps_{0}}{\wsigma+g_{\sigma}}\simeq \int \ddp \eps_{0}\rho_{\sigma}\label{Energys}
\end{equation}
is valid for $p\in [-\pi,\pi]$. 
Finally we consider the fermion density:
\begin{equation}
2\pi n_{\sigma}= \int_{-\pi}^{\pi}\ddp\frac{\nu_{\sigma}(p)}{\nu(p)}.\label{nsigma}\end{equation}
 As in the case of eq. (\ref{nuinvapp}), the integrand $\nu_{\sigma}/\nu$ can be replaced by
\begin{eqnarray}
\frac{\nu_{\sigma}}{\nu}&\simeq &\frac{\nu_{\sigma}}{\sigma-1+\nu_{\sigma}}-\sum_{\sigma'=\sigma+1}^{K}\left(\frac{1}{\sigma'-1}-\frac{1}{\sigma'-1+\nu_{\sigma'}}\right)-\left(\frac{1}{K}-\frac{1}{K+\nu_{\B}}\right)\nonumber\\
&=&G_{\sigma}\left(\wsigma+g_{\sigma}\right)^{-1}-\sum_{\sigma'=\sigma+1}^{K}G_{\sigma'}^2 \left(\wFsp +g_{\sigma'}\right)^{-1}-G_{\rm c}^2\left(\wB+g_{\rm c}\right)^{-1}\nonumber\\
&\simeq&G_{\sigma}\rho_{\sigma}-\sum_{\sigma'=\sigma+1}^{K}G_{\sigma'}^{2}\rho_{\sigma'}-G_{\rm c}^{2}\rho_{\rm c}\quad \mbox{for }2\le \sigma \le K,\label{nusimga}
\end{eqnarray}
\begin{equation}
\nu_{1}/\nu\simeq 1-\sum_{\sigma=2}^{K}G_{\sigma}^2 \rho_{\sigma}-G_{\rm c}^{2}\rho_{\rm c}.\label{nu1}
\end{equation}

Combining eqs. (\ref{Etot}), (\ref{Energyc}) and (\ref{Energys})$-$(\ref{nu1}), we obtain
\begin{eqnarray}
2&\pi& \left(E_{\rm tot}/N-\sum_{\sigma=1}^{K}\zeta_{\sigma}n_{\sigma} +\zeta_{1}\right)\nonumber\\&\simeq& \frac{2\pi^3}{3}+\int_{-\pi}^{\pi}\ddp \eps_{0}\left[\exp\left(\frac{\eps_{0}}{T}\right)-1\right]^{-1}+G_{\rm c}\int_{-\pi}^{\pi}\ddp \eps_{\rm c}\rho_{\rm c}+\sum_{\sigma=2}^{K}G_{\sigma}\int_{-\pi}^{\pi}\ddp\eps_{\sigma}\rho_{\sigma}, \label{Ezeta}
\end{eqnarray}
which is valid at low temperatures. The third term in the right-hand side is the contribution of g-ons with energy $\eps_{\rm c}$, the momentum scaling factor $G_{\rm c}$, and the statistical parameter $g_{\rm c}$. The fourth term can also be regarded as the energy of other g-ons for the spin degrees of freedom. 

Similarly, the entropy is expressed by the sum of g-on contributions. First we study the boson part of the entropy
\begin{equation}
2\pi s_{\rm c}\equiv \int_{-\pi}^{\pi}\frac{\ddp}{\nu}\left[\left(1+\nu_{\B}\right)\ln\left(1+\nu_{\B}\right)-\nu_{\B}\ln\nu_{\B}\right]. \label{sc}
\end{equation}
Since $\nu$ is written as
\begin{equation}
\nu=\wB^{-1}+\sum_{\sigma=1}^{K}\left[1+\left(1+\wB\right)\exp\left(-\zeta_{\sigma}/T\right)\right]^{-1}, 
\end{equation}
we can replace $\nu$ by $\wB^{-1}+K$ at low temperatures in the rapidity region $\Pic$. 
On the other hand, the contribution from the outside of $\Pic$ is negligible (see Appendix D). Hence we can rewrite the integrand of the entropy (\ref{sc}) as 
\begin{eqnarray}
& &\frac{\left(1+\wB\right)\ln\left(1+\wB\right)-\wB \ln\wB}{1+K \wB}\nonumber\\
&\simeq&\frac{\left(1+\wc\right)\ln\left(1+\wc\right)-\wc \ln\wc}{1+K \wc}\nonumber\\
&=&G_{\rm c}\left[\left(\rhoc+\rhoc^{*}\right)\ln\left(\rhoc+\rhoc^*\right)-\rhoc\ln\rhoc-\rhoc^*\ln\rhoc^* \right], \label{scrho}
\end{eqnarray}
where $\rho^{*}_{\rm c}\equiv 1-g_{\rm c}\rho_{\rm c}$. 
The relation (\ref{scrho}) is valid even outside of $\Pic$ where both $\wB$ and $\wc$ are much larger than unity and hence the contribution from the region is negligible. Thus we obtain the low temperature expression for $s_{\rm c}$ as
\begin{equation}
2\pi s_{\rm c}\simeq G_{\rm c}\int_{-\pi}^{\pi}\ddp\left[\left(\rhoc+\rhoc^*\right)\ln\left(\rhoc +\rhoc^*\right)-\rhoc \ln \rhoc-\rhoc^*\ln\rhoc^*\right]. \label{sc2}
\end{equation}
Similarly, we can rewrite the fermion part of entropy as
\begin{eqnarray}
2\pi s_{1}&\equiv&-\int_{-\pi}^{\pi}\frac{\ddp}{\nu}\left[\left(1-\nu_{1}\right)\ln\left(1-\nu_{1}\right)+\nu_{1}\ln\nu_{1}\right] \nonumber\\
&\simeq&-\frac{\partial }{\partial T}\int_{-\pi}^{\pi}\ddp T\ln\left[1-\exp\left(-\frac{\eps_{0}}{T}\right)\right]. \label{s1}
\end{eqnarray}
and
\begin{eqnarray}
2\pi s_{\sigma}&\equiv& -\int_{-\pi}^{\pi}\frac{\ddp}{\nu}\left[\left(1-\nu_{\sigma}\right)\ln\left(1-\nu_{\sigma}\right)+\nu_{\sigma}\ln\nu_{\sigma}\right]\\
&\simeq&G_{\sigma}\int_{-\pi}^{\pi}\ddp\left[\left(\rho_{\sigma}+\rho_{\sigma}^{*}\right)\ln\left(\rho_{\sigma}+\rho_{\sigma}^*\right)-\rho_{\sigma}\ln\rho_{\sigma}-\rho_{\sigma}^*\ln\rho_{\sigma}^*\right]\quad \mbox{for }2\le \sigma \le K,\label{ssigma}  
\end{eqnarray}
where $\rho^{*}_{\sigma}\equiv 1-g_{\sigma} \rho_{\sigma}$. 
With these preliminaries we can derive the thermodynamic potential at low temperatures. From eqs. (\ref{Ezeta}), (\ref{s1}), and (\ref{ssigma}), we obtain
\begin{eqnarray}
2&\pi&\left(E_{\rm tot}/N -\sum_{\sigma=1}^{K}\zeta_{\sigma}n_{\sigma}-Ts_{\rm tot}+\zeta_{1}\right)\nonumber\\
&\simeq& 2\pi^3/3+T\int_{-\pi}^{\pi}\ddp \ln\left(1-e^{-\eps_{0}/T}\right)-TG_{\rm c}\int_{-\pi}^{\pi}\ddp \ln\left(1+\wc^{-1}\right)-T\sum_{\sigma=2}^{K}G_{\sigma}\int_{-\pi}^{\pi}\ddp \ln\left(1+\wsigma^{-1}\right).\nonumber\\
\label{Omegasigma}
\end{eqnarray}
The second term in the right-hand side of (\ref{Omegasigma}) is the same as $\Omega_{\rm ph}$ (eq. (\ref{omegaph})). Subtracting $\Omega_{\rm ph}$, we obtain the compact expression for the thermodynamic potential $\Omega_{\it t\mbox{-}J}$ as
\begin{equation}
2\pi \left(\Omega_{\it t\mbox{-}J}/N+\zeta_{1}\right)\simeq \pi^3/3
-TG_{\rm c}\int_{-\pi}^{\pi}\ddp \ln\left(1+\wc^{-1}\right)-T\sum_{\sigma=2}^{K}G_{\sigma}\int_{-\pi}^{\pi}\ddp \ln\left(1+\wsigma^{-1}\right).\label{Omegaunpo}
\end{equation}
The expression (\ref{Omegaunpo}) is a main result in this subsection. It reproduces, as it should, the result in subsection 4-2 in the SU(2,1) case. 
\subsection{Unpolarized case}
In the unpolarized case, we set $\zeta_{\sigma}=\zeta$ and $\wFs=\wF$ for $1\le \sigma \le K$. Then the low temperature region is defined by
\begin{equation}
\exp\left(\zeta/T\right)\gg 1.\label{lowdef} 
\end{equation}
The main purpose in this subsection is to derive the expression (\ref{Omegamulti}). 

In terms of $\wB$ and $\wF$, equation (\ref{epsX}) is rewritten as
\begin{equation}
\exp\left(\eps_{0}/T\right)=\left(1+\wB^{-1}\right)\left(1+\wF^{-1}\right)^{K}\label{wFwB}.
\end{equation} 
The boson contribution is the same as in the previous case; we can replace $\left(\wB+g_{\rm c}\right)^{-1}$ by $\rho_{\rm c}=\left(\wc+g_{\rm c}\right)^{-1}$ for $p\in\left[-\pi,\pi\right]$. 

On the other hand, the fermion part is different from the polarized one. 
With the use of the relation $\wB=\wF \exp\left(\zeta/T\right)-1$, equation (\ref{wFwB}) is rewritten as
\begin{equation}
\exp\left(\eps_{0}/T\right)=\left[1-\wF^{-1} \exp\left(-\zeta/T\right)\right]^{-1}\left(1+\wF^{-1}\right)^{K}. \label{wFwB2}
\end{equation}
In the rapidity region $\PiF$ where $\wF^{-1}\exp\left(-\zeta/T\right)$ is much smaller than unity, the solution of equation (\ref{wFwB2}) is given by
\begin{equation}
\wF \simeq \left[\exp\left(\eps_{0}/KT\right)-1\right]^{-1}\quad \mbox{for }p\in \PiF. \label{wF}
\end{equation}
With the use of eqs. (\ref{wBrhoc}) and (\ref{wF}), we rewrite $E_{\rm tot}$ (\ref{Etot}). In this case, the integrand in (\ref{Etot}) is replaced by
\begin{eqnarray}
\frac{2\pi E_{\tot}}{N}&=&\int_{-\pi}^{\pi}\ddp
\frac{\eps_{0}}{\nu}\simeq\int_{-\pi}^{\pi}\ddp \eps_{0}\left[\frac{1}{K \nu_{\rm F}}-\left(\frac{1}{K}-\frac{1}{K+\nu_{\B}}\right)\right]\nonumber\\
&\simeq&\frac{2\pi^3}{3K}+\int_{-\pi}^{\pi}\ddp\left(\frac{\eps_{0}}{K}\right)\left[\exp\left(\frac{\eps_{0}}{KT}\right)-1\right]^{-1}-G_{\rm c}^2\int_{-\pi}^{\pi}\ddp \eps_{0}\rho_{\rm c}. \label{Energy2}
\end{eqnarray}
Similarly, the electron density $n_{\rm e}$ is rewritten as 
\begin{eqnarray}
2\pi n_{\rm e}&=&K\int_{-\pi}^{\pi}\ddp\nu_{\rm F}/\nu\nonumber\\
&=&2\pi -G_{\rm c}\int_{-\pi}^{\pi}\ddp \left(\wB +g_{\rm c}\right)^{-1}\nonumber\\
&\simeq& 2\pi -G_{\rm c}\int_{-\pi}^{\pi}\ddp \rho_{\rm c}. \label{ne2}
\end{eqnarray}
The entropy is given by
\begin{equation}
2\pi s_{\tot}\simeq 2\pi s_{\rm c}-\frac{\partial}{\partial T}\int_{-\pi}^{\pi}\ddp T\ln\left[1-\exp\left(-\frac{\eps_{0}}{KT}\right)\right], \label{stot2}
\end{equation}
where $s_{\rm c}$ is given by eq. (\ref{sc2}). The second term in the right-hand side in eq. (\ref{stot2}), which comes from the fermion part, is the same as the entropy of free boson with the energy $\eps_{0}/K$. With the use of eqs. (\ref{Omegatotph}), (\ref{Energy2}), (\ref{ne2}), and (\ref{stot2}), we obtain the low temperature expression for the thermodynamic potential:
\begin{eqnarray}
2&\pi& \left(\frac{\Omega_{\it t\mbox{-}J}}{N}+\zeta\right)\simeq\frac{\pi^3}{3}\left(\frac{2}{K}-1\right)-TG_{\rm c}\int_{-\pi}^{\pi}\ddp \ln\left(1+\wc^{-1}\right)\nonumber\\&-&T\int_{-\pi}^{\pi}\ddp \ln\left[1+e^{-\eps_{0}/KT}+e^{-2\eps_{0}/KT}+\cdots+e^{-\left(K-1\right)\eps_{0}/KT}\right]. \label{Omegamulti}
\end{eqnarray} 
The second term in the right-hand side of eq. (\ref{Omegamulti}) comes from the boson part and is the same as in the polarized case. The third term, which comes from the fermion part, is the thermodynamic potential of {\it parafermions} with order $K$, energy $\eps_{0}/K$. 

The low temperature condition (\ref{lowdef}) is satisfied at all temperatures for the SU($K$) Haldane-Shastry model, where $\zeta \rightarrow \infty$. Hence the SU($K$) Haldane-Shastry model in the absence of external field is thermodynamically equivalent to free parafermions {\it at all temperatures}. 
\section{Discussion}
In $\S$ 2, we constructed the thermodynamics for SU($K$,$1$) {\it t\mbox{-}J}  model. Here we compare two earlier methods \cite{Wang,squeeze,Haldane91} with the method in $\S$ 2. 

One is the method taken by Haldane\cite{Haldane91} and Wang et al. \cite{Wang} It is based on the empirical rule obtained by numerical diagonalization of small systems. When we set $h=0$, the results in \S 4 reproduce those obtained in ref. (\citen{Wang}) as we show in Appendix E. In another limit $\zeta \rightarrow \infty$, we reproduce the results for the Haldane-Shastry model given in ref. (\citen{Haldane91}).  In this approach, thermodynamics is given in terms of spinons and holons and hence the low temperature g-on description is easily obtained.  The trouble with their approach is that it is hard to justify microscopically. In principle, we can refer the issue to the Yangian representation theory.  In the practical sense, however, this approach has not turned out useful in the case of SU($K$,$1$) ($K\ge 3$) case, since the state-counting is too complicated to make an empirical rule for the energy spectrum. \cite{Haltani} 

Another method has been proposed by Ha and Haldane. \cite{squeeze} They adopted a string hypothesis for the SU($K$) Haldane-Shastry model.  The string hypothesis is applicable to the SU($K$,1) model with a slight modification. However it is not straightforward to obtain the low temperature description in terms of free g-ons and parafermions.

Finally the method \cite{S2} we adopted in section 2 provides the easiest way to calculate the thermodynamic quantities and to obtain the low temperature description.  However it is hard to see the thermodynamical equivalence of SU($K$, $1$) {\it t\mbox{-}J}  model with the free fractional particles {\it at all temperatures}. It has been proven for the special case $K=2$ in section 3.  

In the last section, we presented the low temperature description of the thermodynamics of {\it t-J}  model and its multi-component generalization. When all chemical potentials are different from each other, the g-on description accounts for the thermodynamics. On the other hand, we must invoke free parafermion in order to describe the thermodynamics in the unpolarized case. The result in section 6 suggests that a framework unifying the parafermion and fractional exclusion statistics should be worth building.
\section{Summary}

In this paper, we constructed exact thermodynamics of the supersymmetric SU($K$,1) {\it t-J}  models (\ref{lattice}) and discussed the relation between the models and the fractional exclusion statistics. 
The thermodynamic potential for the {\it t-J} model (\ref{lattice}) is given in a simple form. In the SU(2,1) case, the exact thermodynamic potential is reproduced in terms of free spinons and holons obeying the fractional exclusion statistics. At low temperatures, the statistical interaction between spinons and holons does not work and the spin-charge separation occurs. Both spinons and holons obeying the g-on statistics in the temperature region. These properties can be seen in the numerical results on the spin and charge susceptibilities and specific heat. For the SU($K$,1) ( $K \ge 3$ ) case, the low temperature properties are described in terms of g-ons or parafermions, both of which are free particles. 
\section*{Acknowledgement}
We acknowledge the support by Grant-in-Aid from the Ministry of Education, Science, and Culture, Japan.
\appendix
\section{Explicit Form for $\mu_{\B}$}
In this appendix, we derive the explicit form for $\mu_{\B}$.
Taking the logarithm of eq. (\ref{Xdef}), we obtain
\begin{equation}
\epsilon=\mu_{\B}-T\ln X.\label{eps0}
\end{equation}
On the other hand, equation (\ref{epsilondef}) gives
\begin{eqnarray}
\epsilon(p=0)=\epsilon(k=0)&=&\int_{0}^{\infty}\dk' \nu (k')k'\nonumber\\
&=&\int_{0}^{\pi}\ddp k(p)\nonumber\\
&=&\int_{0}^{\pi}\ddp\int_{0}^{p}\ddp'\frac{\partial k\left(p'\right)}{\partial p'}\nonumber\\
&=&\int_{0}^{\pi}\ddp\int_{0}^{p}\ddp'\frac{1}{\nu(p')}\nonumber\\
&=&\int_{0}^{\pi}\ddp
\frac{\pi-p}{\nu(p)}\label{ep0}.
\end{eqnarray}
Combining eqs. (\ref{eps0}) and (\ref{ep0}), the expression for $\mu_{\B}$ is obtained as eq. (\ref{muB}).
\section{Derivation of Eq. (\protect\ref{MATH})}
In this appendix, we derive the relation (\ref{MATH}). First we prove the following relation:
\begin{equation}
\frac{\partial \ln X(p)}{\partial p}=-\frac{p}{T\nu(p)}.\label{relation}
\end{equation}
Differentiating both sides of $(\ref{epsX})$ with respect to $p$, we obtain
\begin{eqnarray}
-p&=&T\left[\sum_{\sigma=1}^{K}\frac{1}{\exp(-\zeta_{\sigma}/T)+X }
+\frac{1}{1-X}\right]\frac{\partial X}{\partial p}\nonumber\\
&=&\frac{T\nu}{X}\frac{\partial X}{\partial p}\nonumber\\
&=&T\nu\frac{\partial \ln X}{\partial p}\label{derive}.
\end{eqnarray}
By dividing  (\ref{derive}) by $\nu$, we obtain the relation (\ref{relation}). 
Next we integrate partially the left-hand side of eq. (\ref{MATH}).
\begin{eqnarray}
\int_{0}^{\pi}\ddp\ln\left[\frac{X(p)}{X(0)}\right]&=&-\int_{0}^{\pi}\ddp\left(\pi -p\right)'\ln\left[\frac{X(p)}{X(0)}\right]\nonumber\\
&=&\int_{0}^{\pi}\ddp\left(\pi-p\right)\frac{\partial \ln X(p)}{\partial p}\nonumber\\
&=&-\int_{0}^{\pi}\ddp\frac{\left(\pi -p\right)p}{T \nu}\label{partialint}.
\end{eqnarray}
From the second to the third lines, we have used the relation (\ref{relation}). 
\section{Calculation of \protect$s_{\tot}$}
In this appendix, we derive the expression ( \protect\ref{Stot}). 
First, we rewrite eq. (\ref{entropy}) as
\begin{equation}
 2\pi s_{\tot}=
\int_{-\pi}^{\pi} \frac{\ddp}{\nu(p)}\left\{\ln\left[\frac{1+\nu_{\B}}{\prod_{\sigma=1}^{K}(1-\nu_{\sigma})}\right]+\sum_{\sigma=1}^{K}\nu_{\sigma}\ln\left(\nu_{\sigma
}^{^-1}-1\right)+\nu_{\B}\ln(\nu_{\B}^{-1}+1)\right\}\label{Stotint}
\end{equation}
From eq. (\ref{pe}), we know that
\begin{equation}
\ln\left[\frac{1+\nu_{\B}}{\prod_{\sigma=1}^{K}(1-\nu_{\sigma})}\right]=\frac{\eps_{0}}{T}.\label{B1}
\end{equation}
With the use of eq. (\ref{Xdef}), we find that
\begin{equation}
\ln \left(\nu^{-1}_{\sigma}-1\right)=-\zeta_{\sigma}/T-\ln X, \quad \ln (\nu_{\B}^{-1}+1)=-\ln X\label{B2}. 
\end{equation}
By using eqs. (\ref{B1}) and (\ref{B2}), we obtain
\begin{eqnarray}
2\pi s_{\tot}&=&\frac{1}{T}\int_{-\pi} ^{\pi}\ddp\frac{\eps_{0}}{\nu}\nonumber\\
&&-\frac{1}{T}\int_{-\pi}^{\pi} \ddp\frac{\sum_{\sigma=1}^{K}\zeta_{\sigma}\nu_{\sigma}}{\nu}-\int_{-\pi}^{\pi} \ddp \frac{(\sum_{\sigma=1}^{K}\nu_{\sigma}+\nu_{\B})\ln X}{\nu}\label{B3}\\
&=&\frac{1}{ T}\int_{-\pi} ^{\pi}\ddp\frac{\eps_0}{\nu}-\int_{-\pi}^{\pi}\ddp\ln X(p) -2\pi(\sum_{\sigma=1}^{K}\zeta_{\sigma} n_{\sigma})/T. \label{B4}
\end{eqnarray}
In going from eqs. (\ref{B3}) to (\ref{B4}), we have used eq. (\ref{ne}).
\section{Contribution to the Entropy from Outside of $\Pi_{\rm c}$ }
In this appendix, we show that the contribution from the outside of the rapidity region $\Pic$ to the boson part in entropy eq. (\ref{sc}) is negligible. 

Since $\nu_{\rm B}$ is much smaller than unity outside $\Pi_{\rm c}$, the numerator in eq. (\ref{sc}) is rewritten as 
\begin{equation}
\left[\left(1+\nu_{\rm B}\right)\ln\left(1+\nu_{\rm B}\right)-\nu_{\rm B}\ln\nu_{\rm B}\right]\simeq \wB^{-1}\ln\wB.
\end{equation}
Except the neighborhood of $p=p_{1}=\pi$, the distribution function $\nu_{1}$ is order of unity and hence the integrand in eq. (\ref{sc}) is much smaller than unity. In the neighborhood of $p=\pi$, which is nothing but the region $\Pi_{1}$, $\nu_{1}$ has the form
\begin{equation}
\nu_{1}=\left(\wFl +1\right)^{-1}\simeq =1-\exp\left(-\eps_{0}/T\right).
\end{equation}
With the use of the relation $\wB =\wFl \exp\left(\zeta_{1}/T\right)-1$, we obtain 
\begin{eqnarray}
\left(\wB \nu\right)^{-1}\ln\wB&\simeq &\left(\ln \wFl +\zeta_{1}/T\right)\exp\left(-\zeta_{1}/T\right)
\nonumber\\
&\simeq&\left\{\zeta_{1}/T-\ln\left[\exp\left(\eps_{0}/T\right)-1\right]\right\}\exp\left(-\zeta_{1}/T\right)\ll 1.  
\end{eqnarray}
Then we find that the contribution from the region near $p=\pi$ is also negligible. 
\section{Reduction to Wang-Liu-Coleman's Result}
Here we show that our result reduces to that of ref. (\citen{Wang}) in the absence of magnetic field.  Since $\eps_{\up}=\eps_{\dn}\equiv\eps_{\rm s}$ with $h=0$, we set $w_{\up}=w_{\dn}\equiv w_{s}$ from eq. (\ref{therm}). In this case, equation (\ref{therm}) is written as
\begin{equation}
\eps_{\rm s}/T=\ln w_{\rm s}-\frac12 \ln(1+w^{-1}_{\rm h}),\label{Wang1}
\end{equation}
\begin{equation}
\eps_{\rm h}/T=\ln\left(1+w_{\rm h}\right)+\ln(1+w^{-1}_{\rm s})-\frac12 \ln(1+w^{-1}_{\rm h}).\label{Wang2}
\end{equation}
The thermodynamic potential eq. (\ref{OmegaFES}) becomes
\begin{equation}
2\pi\left(\Omega/N+\zeta \right)=-T\int_{-\pi}^{\pi}\ddp\left[\ln\left(1+w^{-1}_{\rm s}\right)+\frac12 \ln\left(1+w^{-1}_{\rm h}\right)\right].\label{OmegaWang}
\end{equation}
From eq. (\ref{Wang1}), we obtain
\begin{equation}
\frac12 \ln\left(1+w^{-1}_{\rm h}\right)+\ln\left(1+w^{-1}_{\rm s}\right)=\ln\left(1+w_{\rm s}\right)-\eps_{\rm s}/T.\label{Wang3}
\end{equation}
The substitution (\ref{Wang3}) into (\ref{OmegaWang}) yields
\begin{eqnarray}
2\pi\left(\Omega/N+\zeta\right)&=&-T\int_{-\pi}^{\pi}\ddp \ln\left(1+w_{\rm s}\right)+\int_{-\pi}^{\pi}\ddp \eps_{\rm s}\nonumber\\
&=&\frac{\pi^3}{3}-T\int_{-\pi}^{\pi}\ddp\ln\left(1+w_{\rm s}\right).
\end{eqnarray}
A slight rearrangement of eq. (\ref{Wang1}) gives
\begin{equation}
-T\ln w_{\rm s}=-\eps_{\rm s}-\frac{T}{2}\ln\left(1+w^{-1}_{\rm h}\right).\label{Wang4} 
\end{equation}
Adding both sides of eqs. (\ref{Wang1}) and (\ref{Wang2}), we obtain
\begin{equation}
T\ln w_{\rm h}=\zeta-T\ln\left(1+w_{\rm s}\right). \label{Wang5}
\end{equation}
Identifying $\zeta-\pi^2/6$, $-T\ln w_{\rm s}$, and $T\ln w_{\rm h}$ with $\mu$, $\eps_{\rm s}$, $\eps_{\rm c}$ in ref. (\citen{Wang}), we find that eqs. (\ref{OmegaWang}), (\ref{Wang4}), and (\ref{Wang5}) reproduce eqs. (15) and (16) in ref. (\citen{Wang}).

\begin{figure}
\caption{ (a ) $p$ dependence of distribution functions for spinons and holons at zero magnetic field.  Zero temperature and low temperature results are presented schematically. Here the Fermi surface of holon is given by $p_{\rm h}=\sqrt{\pi^2 -4\zeta}$.   The nonexistent region $\vert p\vert \ge \pi$ is also shown in order to emphasize that the spinon distribution function becomes the fermionic distribution function at low temperatures.  (b) $p$ dependence of distribution functions for up spinons and holons in the presence of magnetic field $(h >0)$.  } Here Fermi surface of up spinon is given by $p_{\uparrow}=\sqrt{\pi^2 -4 h}$. 
\label{rho}
\end{figure}
\begin{figure}
\caption{Temperature dependence of the spin susceptibility $\chi_{\rm s}$ for $n_{\rm e}=$1.0 (half-filling), 0.9, 0.8, 0.6,  and 0.4. The crosses on each curve represent the crossover temperature above which the strong spin-charge separation breaks down.  }
\label{chis}
\end{figure}

\begin{figure}
\caption{Temperature dependence of the charge susceptibility $\chi_{\rm c}$ for $n_{\rm e}=$ 0.9, 0.8, 0.6, 0.4, and 0.2. The crosses on each curve represent the crossover temperature above which the strong spin-charge separation breaks down. Peak structure in low temperature regime is due to the divergent density of states at the bottom of the free holon band.}
\label{chic}
\end{figure}

\begin{figure}
\caption{Temperature dependence of the charge susceptibility for $n_{\rm e}=0.8$. The solid curve represents the exact result and dotted one does the contribution of free holons. The cross represents the crossover temperature $T_{\times}$. }
\label{chicne08}
\end{figure}

\begin{figure}
\caption{ (a) Temperature dependence of the specific heat $C$ for $n_{\rm e}=$ 0.9, 0.8, 0.6, and 0.4. The maximum in $C$ around $T \sim 0.7$ is due to spin excitations. The rapid increase in $C$ for $n_{\rm e}$=0.9 and 0.8 below $T \sim 0.1$ is due to charge excitations. 
(b) The blow-up of the low temperature behavior of the specific heat.} 
\label{heat}
\end{figure}

\begin{figure}
\caption{Decomposition of the specific heat into spinon and holon contributions at $n_{\rm e}=0.8$. The solid curve represents the exact result for the {\it t-J}  model.  The dashed-dotted one shows the contribution of free holons, and the dotted one does that of fermionic spinons. The sum of the two contributions to the specific heat is also shown. 
$T_{\times}$ is denoted by $\times$. }
\label{heatne08}
\end{figure}

\begin{figure}
\caption{The schematic profile of $\nu(p)^{-1}$ in the case of $K=2$. The Fermi surfaces $p_{\rm c}$, $p_{\sigma}$ are shown. Near $p=p_{\sigma}$, the rapidity dependence of $\nu(p)^{-1}$ is determined mainly by $\nu_{\sigma}(p)$ while the behavior of $\nu(p)^{-1}$ around $p_{\rm c}$ is determined by $\nu_{\rm B}(p)$.}
\label{nuprofile}
\end{figure}

\begin{thebibliography}{99}
\bibitem{Suth}B. Sutherland: J. Math. Phys. {\bf 12} (1971) 246; {\it ibid.}
{\bf 12} (1971) 251; Phys. Rev. A {\bf 4} (1971) 2019; {\it ibid.} {\bf 5} (1972) 1372.
\bibitem{Haldanefrac}F. D. M. Haldane: Phys. Rev. Lett. {\bf 67} (1991) 937.  
\bibitem{Haldane1}F. D. M. Haldane: Phys. Rev. Lett. {\bf 60} (1988) 635.
\bibitem{Shastry}B. S. Shastry: Phys. Rev. Lett. {\bf 60} (1988) 639.
\bibitem{KY}Y. Kuramoto and H. Yokoyama: Phys. Rev. Lett. {\bf 67} (1991) 1338.
\bibitem{Ha}Z. N. C. Ha and F. D. M. Haldane: Phys. Rev. Lett. {\bf 73} (1994) 2887.
\bibitem{Wang}D. F. Wang, J. T. Liu, P. Coleman: Phys. Rev. B {\bf 46} (1992) 6639.
\bibitem{squeeze}Z. N. C. Ha and F. D. M. Haldane: Phys. Rev. B {\bf 47} (1993) 12459.

\bibitem{Poly2}A. P. Polychronakos: Phys. Rev. Lett. {\bf 70} (1993) 2329.
\bibitem{S2}B. Sutherland and B. S. Shastry: Phys. Rev. Lett. {\bf 71} (1993) 5.
\bibitem{KuramotoKato}Y. Kuramoto and Y. Kato: J. Phys. Soc. Jpn. {\bf 64} (1995); preprint ( To be published in {\it Proceedings of Pacific conference on Condensed Matter, Complex Materials, and Strongly Correlated systems}).
\bibitem{Bernard}D. Bernard, M. Gaudin, F. D. M. Haldane, and V. Pasquier: J. Phys. A {\bf 26} (1993) 5219. 
\bibitem{Kato}Y. Kato and Y. Kuramoto: Phys. Rev. Lett. {\bf 74} (1995) 1222.
\bibitem{Haldane91}F. D. M. Haldane: Phys. Rev. Lett. {\bf 66} (1991) 1529.
\bibitem{Wu}Y.-S. Wu: Phys. Rev. Lett. {\bf 73} (1994) 922.
\bibitem{Nayak}C. Nayak and F. Wilczek: Phys. Rev. Lett. {\bf 73} (1994) 2740.  \bibitem{Murphy}M. V. N. Murphy and R. Shankar: Phys. Rev. Lett. {\bf 72} (1994) 3629.
\bibitem{Usuki}N. Kawakami, T. Usuki, and A. Okiji: Phys. Lett. A {\bf137} (1989) 287.
\bibitem{Ogata}P.-A. Bares, G. Blatter, and M. Ogata: Phys. Rev. B {\bf 44} (1991) 130. 
\bibitem{Haltani}F. D. M. Haldane: in {\it Correlation Effects in Low-Dimensional Electron Systems} (edited by A. Okiji and N. Kawakami,
Springer-Verlag, 1994) p.3.
%
%
\end{thebibliography}
\end{document}